# Numerical Study of Flow and Cavitation in Single-Liquid Injectors with Sharp-Edged and Chamfered Orifice Inlets

Shreyash Gulati[1,*], Sven R. Walsberg[2,3], Timothy W. Nearhood[2], and Jeffrey D. Moore[2]

[1] *Mechanical Engineering, Lafayette College, PA 18042, USA*

[2] *Division of Engineering, Science, and Technology, The Pennsylvania State University, Altoona College, Altoona, PA, 16601, USA*

[3] *Department of Engineering Science and Mechanics, The Pennsylvania State University, University Park, State College, PA 16802, USA*

**Correspondence:* gulatis@lafayette.edu

**Abstract**: In the current work, the investigators used Computational Fluid Dynamics (CFD) to investigate water flow and cavitation in single-liquid injectors with sharp-edged and chamfered orifice inlets. The $k$-$\omega$ Shear Stress Transport (SST) turbulence model was used along with the Zwart-Gerber-Belamri (ZGB) cavitation model and the Volume of Fluid (VOF) method. The CFD predictions for the volumetric flow rate were validated against existing experimental data for the sharp-edged and chamfered orifice inlets. Pressure drops ranging between 0.07 MPa and 0.59 MPa were considered along with orifice length-to-diameter ratios of 2, 5, and 10. Strong cavitation was observed for the sharp-edged orifice inlet cases with a length-to-diameter ratio of 5 and a pressure drop larger than 0.2 MPa. A hydraulic flip was observed for the sharp-edged orifice inlet with a length-to-diameter ratio of 5 and a pressure drop of 0.59 MPa, which was consistent with the findings from the previous experimental study. Cavitation inception was observed for the chamfered orifice inlet cases with a length-to-diameter ratio of 5 and pressure drops larger than 0.2 MPa. For the sharp-edged orifice inlet with orifice length-to-diameter ratios of 2 and 10, a strong cavitation was observed inside the orifice.

## 1. Introduction

The orifice inlet geometry plays an important role in determining the pressure drop through a single-liquid rocket injector (Kiaoulias et al 2019). For repeatability in manufacturing and pressure drop values, sharp-edge and chamfered orifice inlets are used. Typical sharp-edge and chamfered geometries are shown in Fig. 1. Pressure drop through the injector depends strongly on the diameter ratio, which is the ratio of the orifice diameter to the diameter at the inlet of the injector.

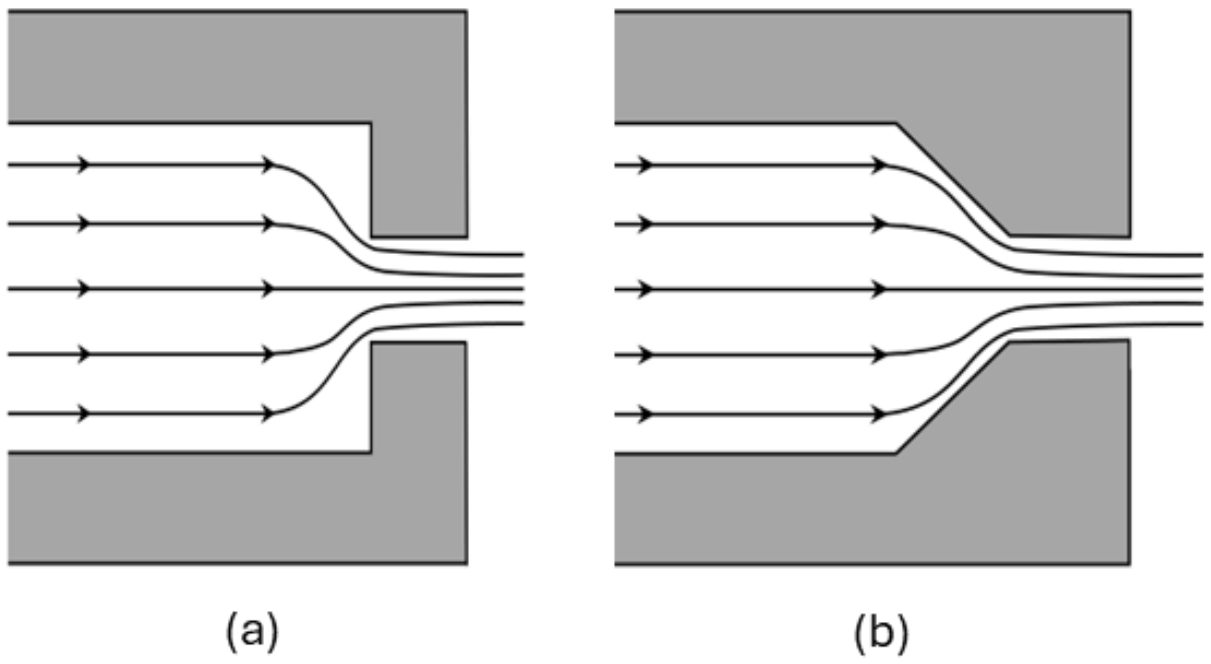

**Fig. 1. Schematic of (a) sharp-edge orifice inlet and (b) chamfered orifice inlet for a liquid injector.**

Cold-flow experiments are generally utilized to test and design liquid rocket injectors. Typically, deionized or distilled water is used in the cold-flow (no combustion) experiments to simulate liquid fuel such as Hydrazine. Reitz and Bracco (1982) utilized cold-flow experiments to study the mechanism of atomization caused by liquid injectors with various sharp-edge and rounded inlet geometries. McHale and Nurick (1974) performed cold-flow experiments to investigate the flow through liquid injectors with non-circular orifices. Kiaoulias et al (2019) carried out cold-flow experiments ($1{,}000 \leq Re \leq 10{,}000$) of single liquid injectors with sharp-edge and chamfered inlets. Here $Re$ is the Reynolds number defined as

$$Re = \frac{\rho V D}{\mu} \tag{1}$$

where $\rho$ is the density of water, $V$ is the mean velocity of water, $D$ is the diameter of the orifice, and $\mu$ is the dynamic viscosity of water. Kiaoulias et al (2019) studied injectors with diameter ratios of $0.16$, $0.4$, and $0.52$, and found large pressure drops across the injector for a sharp-edge inlet specifically for a small orifice with a diameter ratio of $0.16$ and small $L/D$ of $2$ and $5$, where $L$ is the length of the orifice.

Flow through the injector with a sharp-edged orifice inlet, shown in Fig. 1a, can simply be interpreted as flow through a sudden contraction. Several theoretical, empirical, and CFD studies of sudden contraction in pipes have been conducted in the past. Benedict et al (1966) developed compressible-flow solutions to predict the loss coefficient for a sudden contraction in a pipe for turbulent flow and validated the same against experiments for diameter ratios ranging from 0 to 1. Durst and Loy (1985) performed CFD studies to investigate laminar flow through a sudden contraction in a pipe ($23 \leq Re \leq 1213$) with a diameter ratio of $0.53$ and validated the CFD results against experiments. Bullen et al (1987) developed improved theoretical predictions of the loss coefficients for turbulent flow through sudden contractions ($40{,}000 \leq Re \leq 200{,}000$) for diameter ratios ranging from $0$ to $1$. Their predictions compared closely with experimental data. Some of the researchers (Roul and Dash 2011, Kaushik et al 2012) utilized Reynolds Averaged Navier-Stokes (RANS) based models to predict turbulent flow in sudden contractions. Roul and Dash (2011) utilized CFD to study single and multi-phase flow through a sudden contraction (diameter ratio of $0.53$) in a pipe for laminar and turbulent flow ($1{,}000 \leq Re \leq 12{,}000$). They modeled turbulent flow by utilizing the realizable $k$-$\varepsilon$ model. Kaushik et al (2012) studied turbulent, multiphase flow through a sudden contraction in a pipe numerically by utilizing the standard $k$-$\varepsilon$ turbulence model. The diameter ratio in their study was $0.5$.

Injectors with sharp-edge orifice inlets operating at higher pressures are susceptible to cavitation and hydraulic flip (Chew 1973, Nurick 1976, Lin and Katz 1987, Schmidt 1997, Bunnell et al 1999, Tafreshi and Pourdeyhimi 2004, Jung et al 2006, Sou et al 2007, Dabiri et al 2007, Park et al 2008). At higher operating pressures, the liquid turns around the sharp edge at a high velocity, leading to a sudden drop in pressure. When the pressure drops below the saturation pressure, the liquid changes into vapor, and vapor bubbles are formed. These vapor bubbles are pushed out of the injector and depending upon the L/D ratio and the operating pressure of the injector, some of the bubbles may transform back into liquid before they leave the injector. However, for small L/D ratios and higher operating pressures, researchers (Jung et al 2006, Dabiri et al 2007, and Park et al 2008) have observed that the bubbles remain in the vapor state throughout the entire length of the orifice (supercavitation). Hydraulic flip occurs when the liquid inside the orifice separates from the wall leading to a reduction in the effective flow area. The ambient fluid can enter the separation zone, which generally extends throughout the length of the orifice (Dabiri et al 2007).

Researchers have investigated turbulent flow and cavitation in liquid injectors numerically by utilizing a variety of CFD techniques. Large Eddy Simulation (LES) has been considered as the most accurate method for predicting turbulent flow in injectors. However, due to the high computational cost associated with LES, a majority of the researchers resorted to RANS-based models, such as the standard $k$-$\varepsilon$, RNG (Re-Normalization Group) $k$-$\varepsilon$, standard $k$-$\omega$, and $k$-$\omega$ SST (Shear Stress Transport) models. The commonly used cavitation models are the Schnerr-Sauer model and the Zwart–Gerber–Belamri (ZGB) model. Table 1 shows a comprehensive list of turbulence and cavitation models used by several researchers. Since the flow is expected to separate around the sharp-edge orifice inlet, the $k$-$\omega$ SST model was chosen to model the turbulent flow in the current study because it has been reported to predict separation flows with high accuracy. The ZGB model was utilized to predict cavitation in the injector. The work by Perkovic et al (2008), Sadegharani et al (2013), Qiu et al (2016), Mithun et al (2018), Bambhania and Patel (2023), Zainab and Syed (2023), Chen et al (2024), and Keller (2024) are relevant to the current work due to a similarity in the injector geometry as compared to the injector geometry considered in this work. However, none of the above-mentioned studies included the exact geometry and pressure drops as those considered in the current work.

**Table 1: Literature review of CFD-based turbulent flow and cavitation studies of injectors.**

| Author | Turbulent Model(s) | Cavitation Model(s) |
|---|---|---|
| Muss et al (2003) | Standard $k$-$\varepsilon$ | N/A |
| Waidmann et al (2006) | Standard $k$-$\varepsilon$ | N/A |
| Aleiferis et al (2007) | Standard $k$-$\varepsilon$ and RNG $k$-$\varepsilon$ | N/A |
| Tucker et al (2008) | LES, Spalart-Allmaras, and $k$-$\omega$ SST | N/A |
| Perkovic et al (2008) | Standard $k$-$\varepsilon$ | Linear cavitation model |
| Giannadakis et al (2008) | Standard $k$-$\varepsilon$ | Discrete Bubble Model |
| Jayashankara and Ganesan (2010) | Standard $k$-$\varepsilon$ and RNG $k$-$\varepsilon$ | N/A |
| Matsumoto et al (2010) | RNG $k$-$\varepsilon$ | N/A |
| Tsohas and Heister (2011) | $k$-$\omega$ SST | N/A |
| Guezennec et al (2012) | LES | N/A |
| Battistoni and Grimaldi (2012) | k-ζ-f | Two-fluid approach |
| Sadegharani et al (2013) | Realizable $k$-$\varepsilon$ | Schnerr-Sauer |
| Chintala and Subramanian (2013) | Standard $k$-$\varepsilon$ | N/A |
| Benzon et al (2015) | RNG $k$-$\varepsilon$, standard $k$-$\omega$, $k$-$\omega$ SST, and transition SST | N/A |
| Tahmasebi et al (2015) | $k$-$\omega$ SST | Barotropic equation of state |
| Drozda et al (2016) | Blended $k$-$\varepsilon$ and $k$-$\omega$ | N/A |
| Qiu et al (2016) | Transition SST | Zwart–Gerber–Belamri |
| Bicer and Sou (2016) | RNG $k$-$\varepsilon$ | Modified Rayleigh |
| Ghorbani et al (2017) | Standard $k$-$\varepsilon$ | Schnerr-Sauer |
| Hellman (2018) | $k$-$\omega$ SST | N/A |
| Mithun et al (2018) | LES | Three-phase model |
| Shahangian (2020) | $k$-$\omega$ SST | Homogeneous relaxation model |
| Kim et al (2021) | RNG $k$-$\varepsilon$ | N/A |
| Payri et al (2022) | LES | N/A |
| Zainab and Syed (2023) | $k$-$\omega$ SST and LES | N/A |
| Dai et al (2023) | Realizable $k$-$\varepsilon$ | Zwart–Gerber–Belamri |
| Bambhania and Patel (2023) | RNG $k$-$\varepsilon$, Realizable $k$-$\varepsilon$, and $k$-$\omega$ SST | Schnerr-Sauer and Zwart–Gerber–Belamri |
| Uppu et al (2024) | Standard $k$-$\omega$ | N/A |
| Keller (2024) | $k$-$\omega$ SST | N/A |
| Chen et al (2024) | Transition SST | Zwart–Gerber–Belamri |
| Johnson et al (2025) | LES | N/A |
| Gioud et al (2025) | LES | N/A |
| Yuan et al (2025) | Standard $k$-$\varepsilon$ | Schnerr-Sauer |
| Villagomez-Moreno et al (2025) | RNG $k$-$\varepsilon$ | Zwart–Gerber–Belamri |
| Li et al (2025) | LES, Lattice Boltzmann method | Discrete Bubble Model |
| Xi et al (2025) | LES | Modified Schnerr-Sauer |

The objective of the current work was to utilize CFD studies to investigate cavitation and turbulent flow for water flowing through the injector with sharp-edged and chamfered orifice inlets for $L/D = 2$, 5, and 10. This work also evaluates the validity of the hypothesis made by Kiaoulias et al (2019) that for a sharp-edged orifice inlet, larger pressure drops ($\Delta p > 0.2$ MPa) lead to a hydraulic flip, especially for $L/D$ ratios of 2 and 5.

In the following sections, we describe CFD studies that were performed to predict the flow rate of water and cavitation in a single-liquid injector with sharp-edged and chamfered orifice inlets for various length-to-diameter ratios and pressure drops. A comparison was made between the flow rates observed in experiments by Kiaoulias et al (2019) and the flow rates predicted by the current CFD simulations. Further, plots of volume fraction of liquid water are presented for all the cases, and they were examined to evaluate the cavitation regimes.

# 2. Methods

## 2.1 Injector Geometry

The geometry of an injector with a sharp-edge orifice inlet studied in the current work is shown in Fig. 2. The overall length of the injector element is 63.5 mm. The injector is made up of two passages. The liquid first passes through a broader down comer with an internal diameter of $D_{DC} = 3.18$ mm. Subsequently, the liquid passes through a narrower orifice with an internal diameter of $D = 0.51$ mm (diameter ratio of 0.16). The length of the orifice, $L$ was varied to numerically simulate $L/D$ ratios of 2, 5, and 10. Those $L/D$ ratios were the same as some of of the $L/D$ ratios considered in previous cold-flow experiments (Kiaoulias et al, 2019).

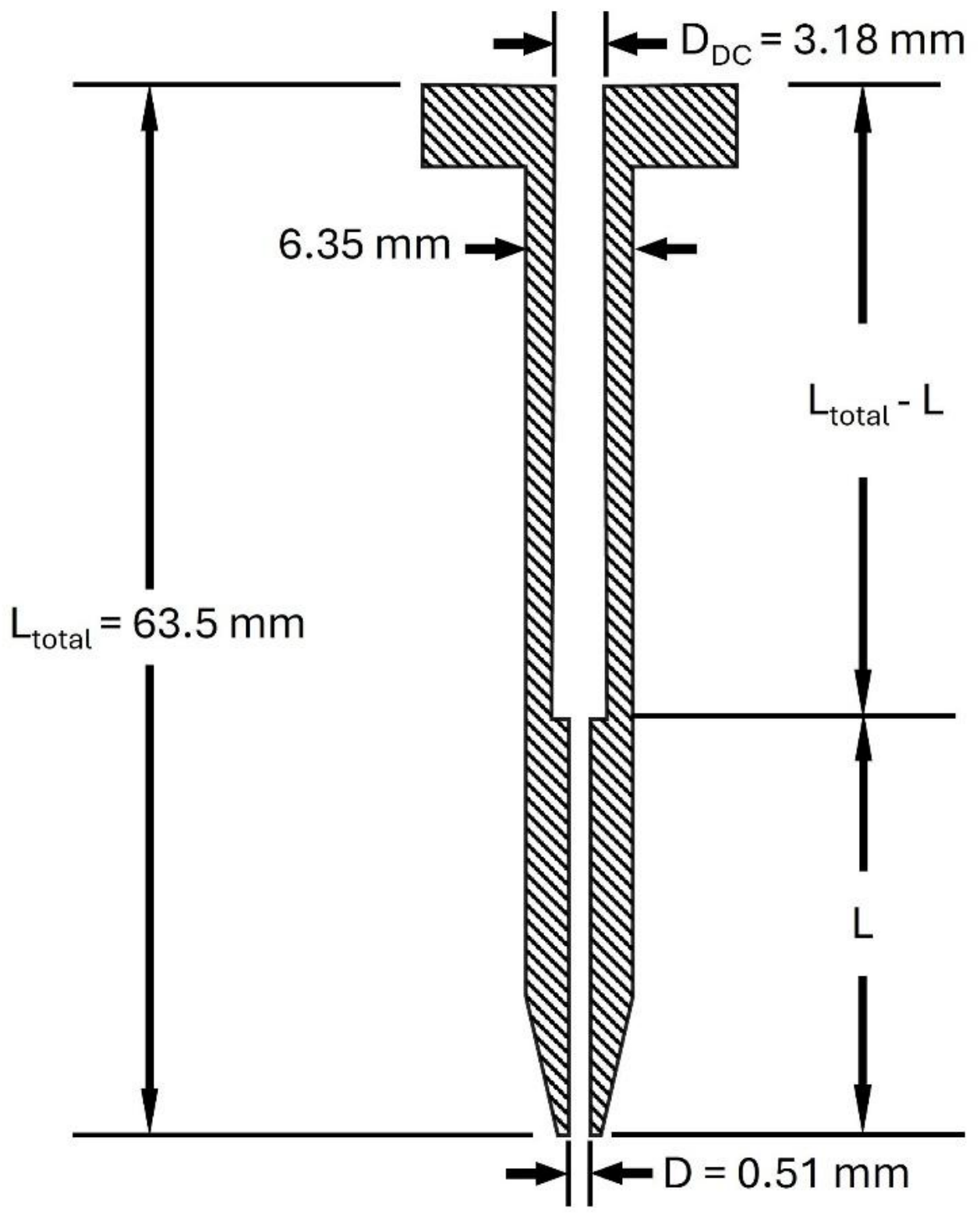

**Fig. 2. Geometry of a liquid injector with a sharp-edged orifice inlet.**

Injectors with a chamfered orifice inlet were also considered. The full cone angle for the chamfered inlet was $120°$. An $L/D$ ratio of 5 was considered for the chamfered orifice inlet cases.

## 2.2. Governing Equations

A two-dimensional axisymmetric fluid domain was considered. Cylindrical coordinates ($r$ , $\theta$ , $z$) were utilized. Then the velocity vector can be written as $\vec{V} = u\hat{e}_r + w\hat{e}_z$, where $u$ is the radial component and $w$ is the axial component of velocity.

There are three phases in the current problem of flow and cavitation in injectors. These are air, liquid water, and water vapor. The Volume of Fluid (VOF) method was used to model the three-phase flow. The volume fraction of the $i^{th}$ phase, $\alpha_i$ was calculated by solving the equation

$$\frac{1}{\rho_i}\left[\frac{\partial}{\partial t}(\alpha_i\rho_i) + \nabla.\left(\alpha_i\rho_i\vec{V_i}\right) = S_{\alpha_i} + \sum_{j=1}^{n}(\dot{m}_{ji} - \dot{m}_{ij})\right], \tag{2}$$

where $\rho_i$ is the density of the $i^{th}$ phase, $\vec{V_i}$ is the velocity vector for the $i^{th}$ phase phase, $S_{\alpha_i}$ is the source term for the $i^{th}$ phase, $n$ is the total number of phases, $\dot{m}_{ji}$ is the mass transfer from the $j^{th}$ phase to the $i^{th}$ phase, and $\dot{m}_{ij}$ is the mass transfer from the $i^{th}$ phase to the $j^{th}$ phase. The volume fractions for all the phases must sum up to 1, i.e.

$$\sum_{i=1}^{n}\alpha_i = 1. \tag{3}$$

Table 2 gives the density and dynamic viscosity of air, liquid water, and water vapor as used in the current study.

**Table 2: Physical properties of fluids at 21°C.**

| Fluid | Density (kg/m$^3$) | Dynamic viscosity (kg/m-s) |
|---|---|---|
| Air | 1.19 | $1.81 \times 10^{-5}$ |
| Liquid water | 998 | $9.74 \times 10^{-4}$ |
| Water vapor | 0.018 | $0.976 \times 10^{-5}$ |

Cavitation was simulated by utilizing the Zwart–Gerber–Belamri model (Zwart et al 2004). The mass transfer from liquid water to water vapor through evaporation is

$$\dot{m}_{lw-wv} = F_{vap} \frac{3\alpha_{nuc}(1-\alpha_{wv})\rho_{wv}}{R_B} \sqrt{\frac{2}{3}\frac{(p_v - p)}{\rho_{lw}}}, \tag{4}$$

where $F_{vap}$ is the evaporation coefficient, $\alpha_{nuc}$ is the nucleation site volume fraction, $\alpha_{wv}$ is the volume fraction of the water vapor, $\rho_{wv}$ is the density of the water vapor, $R_B$ is the vapor bubble radius, $p_v$ is the vapor pressure of water, and $\rho_{lw}$ is the density of liquid water. The mass transfer from water vapor to liquid water through condensation is

$$\dot{m}_{wv-lw} = F_{cond} \frac{3\alpha_{wv}\rho_{wv}}{R_B} \sqrt{\frac{2}{3}\frac{(p - p_v)}{\rho_{lw}}}, \tag{5}$$

where $F_{cond}$ is the condensation coefficient. The vapor bubble radius, $R_B$ was assumed to be equal to $10^{-6}$ m and the nucleation site volume fraction, $\alpha_{nuc} = 5 \times 10^{-4}$. The evaporation coefficient was set to $50$ and the condensation coefficient was set to $0.01$.

The RANS approach was utilized to predict the turbulent flow in the injector. The velocity vector can be written as

$$\vec{V} = \vec{V}_{avg} + \overrightarrow{v'}, \tag{6}$$

where $\vec{V}_{avg}$ is the time-averaged part of the velocity vector and $\overrightarrow{v'}$ is the fluctuating part of the velocity vector. RANS equations treating the three phases as a single liquid in the vector form are

$$\rho \nabla \cdot \vec{V}_{avg} = 0, \tag{7}$$

$$\rho \frac{\partial \vec{V}_{avg}}{\partial t} + \rho \vec{V}_{avg} \cdot \nabla \vec{V}_{avg} + \nabla \cdot \overline{\left(\rho \overrightarrow{v'} \otimes \overrightarrow{v'}\right)} = -\nabla p + \nabla \cdot \mu \left(\nabla \vec{V}_{avg} + \left(\nabla \vec{V}_{avg}\right)^T\right) + \rho \vec{g}, \tag{8}$$

where $\rho = \sum \alpha_i \rho_i$ is the density of the single fluid, $\otimes$ represents the outer vector product, $p$ is the static pressure, $\mu = \sum \alpha_i \mu_i$ is the dynamic viscosity of the single liquid, and $\vec{g}$ is the acceleration due to gravity assumed as $\vec{g} = 9.81\hat{e}_z$ m/s$^2$.

An Eddy Viscosity Model (EVM) (Pope, 2000) is utilized to account for the Reynolds stress $\vec{R} = \overline{v' \otimes v'}$. EVM leads to two additional unknowns- the turbulence kinetic energy $k$ and its specific dissipation rate $\omega$. The $k$-$\omega$ SST model (Menter 1994) is used to close the RANS-EVM setup.

## 2.3 Numerical Modeling

CFD-based models were created by using a commercial software package (ANSYS Fluent). The two-dimensional axisymmetric fluid domain and boundary conditions for the sharp-edge orifice inlet cases are shown in Fig. 3. Similar fluid domain and boundary conditions were used for the chamfered orifice inlet cases. A pressure inlet was specified at the inlet of the downcomer. The outlet of the fluid domain was modeled as a pressure outlet with $p = p_{atm} = 101.325$ kPa. The remaining boundaries were treated as walls with a no-slip boundary condition. The 5 mm length of the domain from the orifice exit to the outlet was chosen to ensure a sufficiently large downstream zone ($\sim 10D$) to minimize the impact of any backflow at the outlet on the numerical prediction of the overall flow.

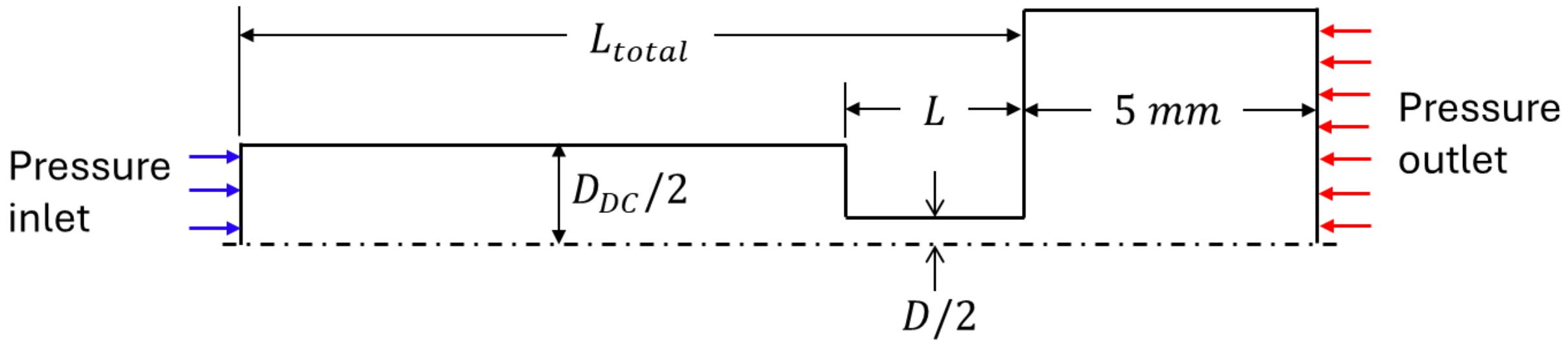

**Fig. 3 Fluid domain and boundary conditions.**

Figs. 4 and 5 show the meshes, which consisted of structured quadrilateral elements in the boundary layer region and unstructured triangular elements elsewhere.

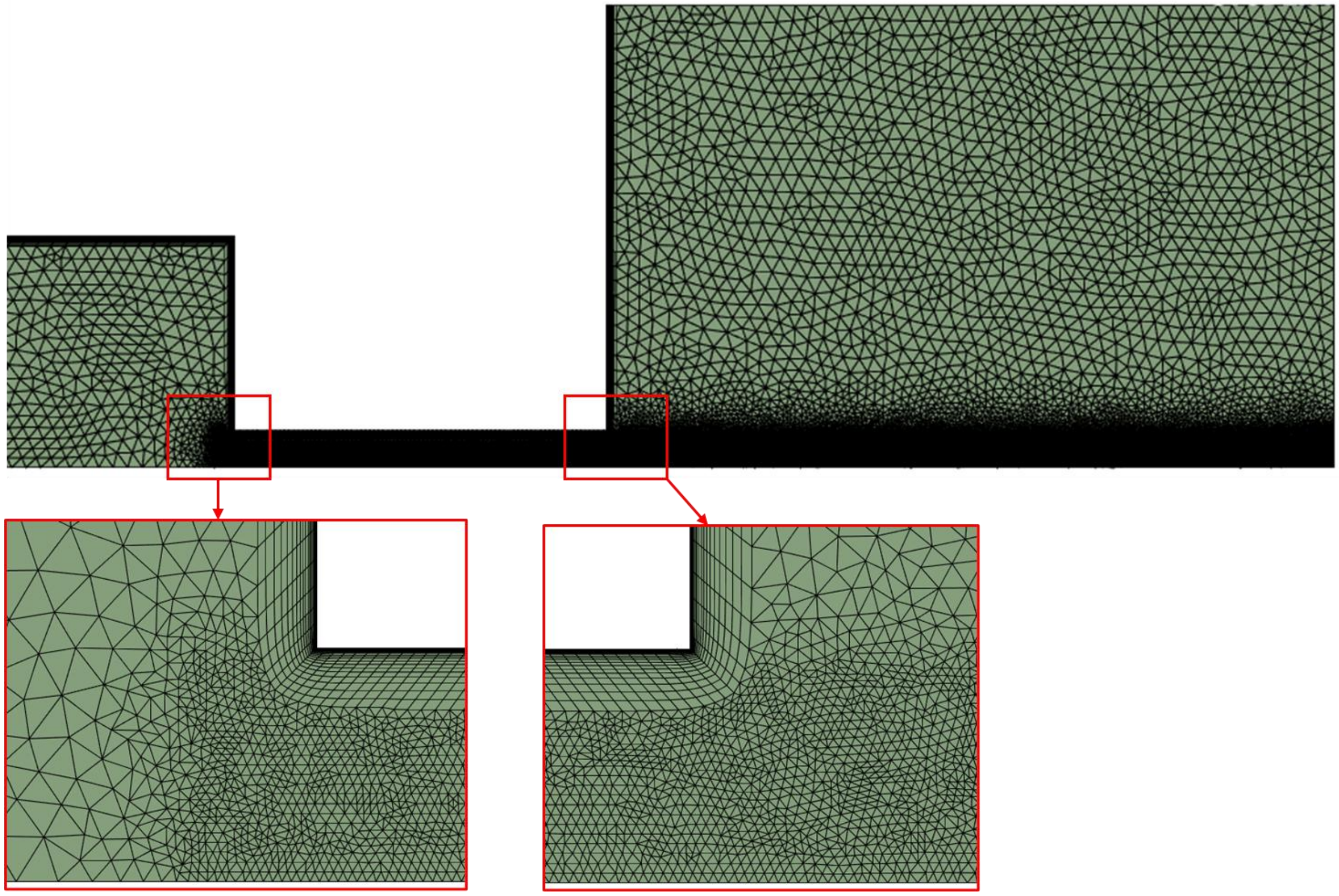

**Fig. 4 Mesh for the sharp-edge orifice inlet with $L/D = 5$.**

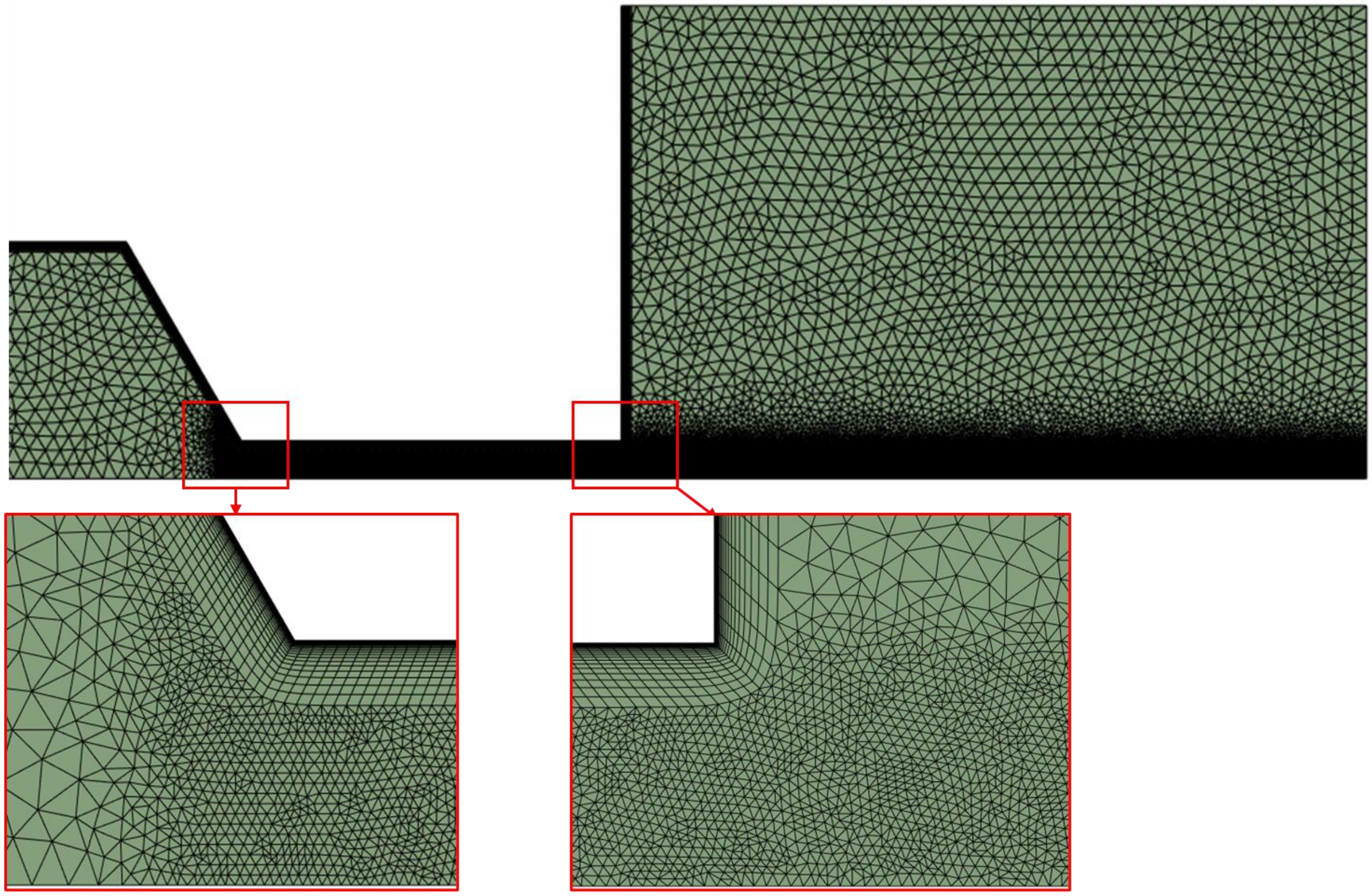

**Fig. 5 Mesh for the chamfered orifice inlet with $L/D = 5$.**

The $y^+$values were kept below 1. The total number of mesh elements for a sharp-edged orifice inlet with $L/D = 5$ was 81,776. The total number of mesh elements for a chamfered orifice inlet with $L/D = 5$ was 86,334. A second-order upwind scheme was used for momentum, volume fraction, vaporization mass, turbulent kinetic energy, and specific rate of dissipation. The Semi-Implicit Method for Pressure Linked Equations-Consistent (SIMPLEC) algorithm was used for the pressure-velocity coupling. The PREssure STaggering Option (PRESTO) scheme was used for the interpolation of pressure. Time integration was performed by utilizing a second-order implicit scheme. A time step of $3 \times 10^{-8}$ s was used. This resulted in a Courant-Friedrichs-Lewy (CFL) number of 0.14 for the sharp-edged orifice inlet with the largest pressure drop of 0.59 MPa as considered in the current study. A mesh refinement study was carried out for the sharp-edge orifice inlet with $L/D = 5$ for $\Delta p = 0.59$ MPa. Table 3 shows the number of mesh elements, volumetric flow rate, and % difference for the mesh refinement study. The % difference shown in Table 3 was calculated by considering the base mesh (81,776 elements) as the reference. The % difference in the flow rate was less than 2% when the mesh was refined from 81,776 elements to 125,109 elements. Therefore, the mesh with 81,776 elements was selected for the current CFD study.

**Table 3: Mesh refinement study for the sharp-edge orifice inlet with $L/D = 5$ for $\Delta p = 0.59$ MPa.**

| Number of Mesh Elements | $Q_{CFD}$(cc/min) | % $Difference$ |
|---|---|---|
| 50,618 | 289.7 | 5.9% |
| 81,776 | 273.5 | |
| 125,109 | 276.5 | 1.1% |

# 3. Results and Discussion

## 3.1 Volumetric flow rates for the sharp-edged and chamfered orifice inlets

Figure 6 shows plots that compare the CFD predicted volumetric flow rate with the experimentally measured volumetric flow rate (Kiaoulias et al 2019) for different pressure drops across the injector for sharp-edged and chamfered orifice inlets with $L/D = 5$. Based upon Fig. 6, the CFD predictions match well with the experimental results for different pressure drops below 0.2 MPa. The CFD model slightly overestimates the volumetric flow rate above 0.2 MPa, with a peak error of 15% for sharp-edged and 9% for chamfered orifice inlets. This may be attributed to the cavitation model for higher flows. It can be concluded that the current CFD models have been validated against experimental measurements by Kiaoulias et al (2019).

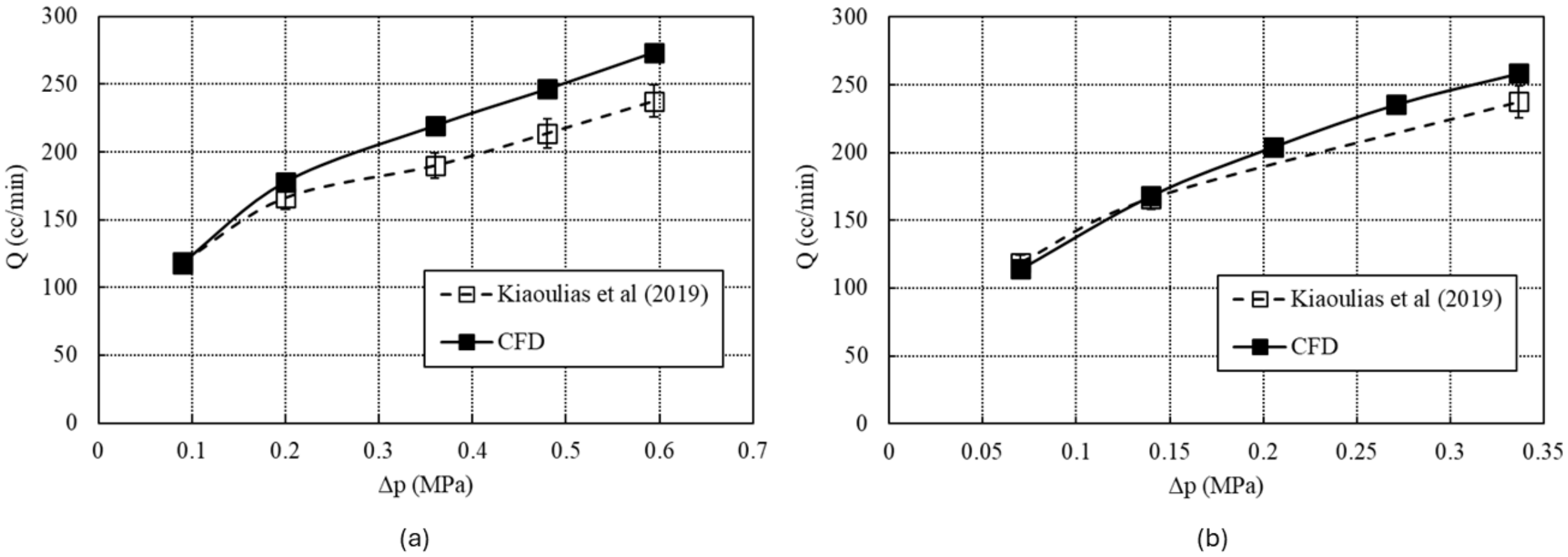

**Fig. 6 Comparison of CFD-predicted and experimentally measured (Kiaoulias et al 2019) volumetric flow rates for different pressure drops for (a) sharp-edge orifice inlet and (b) chamfered orifice inlet with $L/D = 5$.**

## 3.2 Cavitation results for the sharp-edged and chamfered orifice inlets

The plots of volume fraction of the liquid water phase for the sharp-edged orifice inlet cases are presented in Fig. 7. No cavitation was observed from the CFD simulations for $\Delta p = 0.09$ MPa. For $\Delta p = 0.2$ MPa, a cavitation inception was seen close to the sharp edge. As $\Delta p$ was increased to $0.36$ MPa, cavitation became stronger, and the cavitation zone extended to around $80\%$ of the orifice length. For $\Delta p = 0.48$ MPa and $\Delta p = 0.59$ MPa, the cavitation zone extends further downstream inside the orifice, demonstrating very strong cavitation. A hydraulic flip was observed for $\Delta p = 0.59$ MPa. Air can be seen to enter upstream at the orifice exit for that case. Kiaoulias et al (2019) stipulated the presence of a hydraulic flip for all pressure drops larger than $0.2$ MPa. Although the current CFD study did not reveal a hydraulic flip for $\Delta p = 0.36$ MPa and $\Delta p = 0.48$ MPa, the current trend of cavitation becoming stronger with an increase in pressure drop and the presence of a hydraulic flip for $\Delta p = 0.59$ MPa are consistent with the findings from the experimental study by Kiaoulias et al (2019).

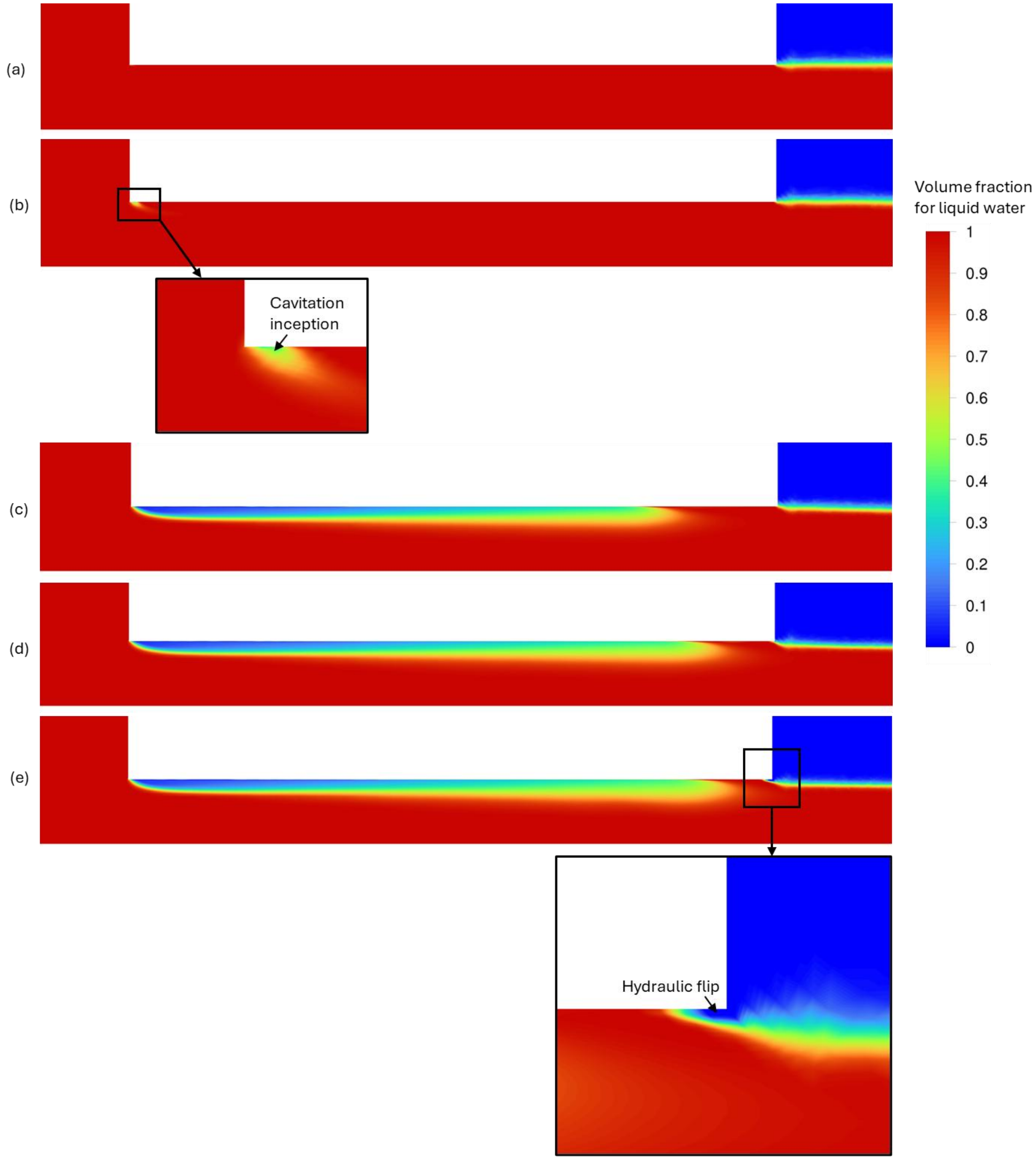

**Fig. 7 Contours of water (liquid) volume fraction for the sharp-edged orifice inlet and $L/D = 5$ and with (a) $\Delta p = 0.09$ MPa (b) $\Delta p = 0.2$ MPa (c) $\Delta p = 0.36$ MPa (d) $\Delta p = 0.48$ MPa, and (e) $\Delta p = 0.59$ MPa.**

Fig. 8 shows the plots of liquid water volume fraction for the chamfered orifice inlet cases. No cavitation was observed for the pressure drops $\Delta p = 0.07$ MPa and 0.14 MPa. However, cavitation inception was seen for the pressure drops $\Delta p = 0.21$ MPa, 0.27 MPa, and 0.34 MPa. Overall, no hydraulic flip was observed for the chamfered orifice inlet cases. This is consistent with the findings from the experimental study by Kiaoulias et al (2019).

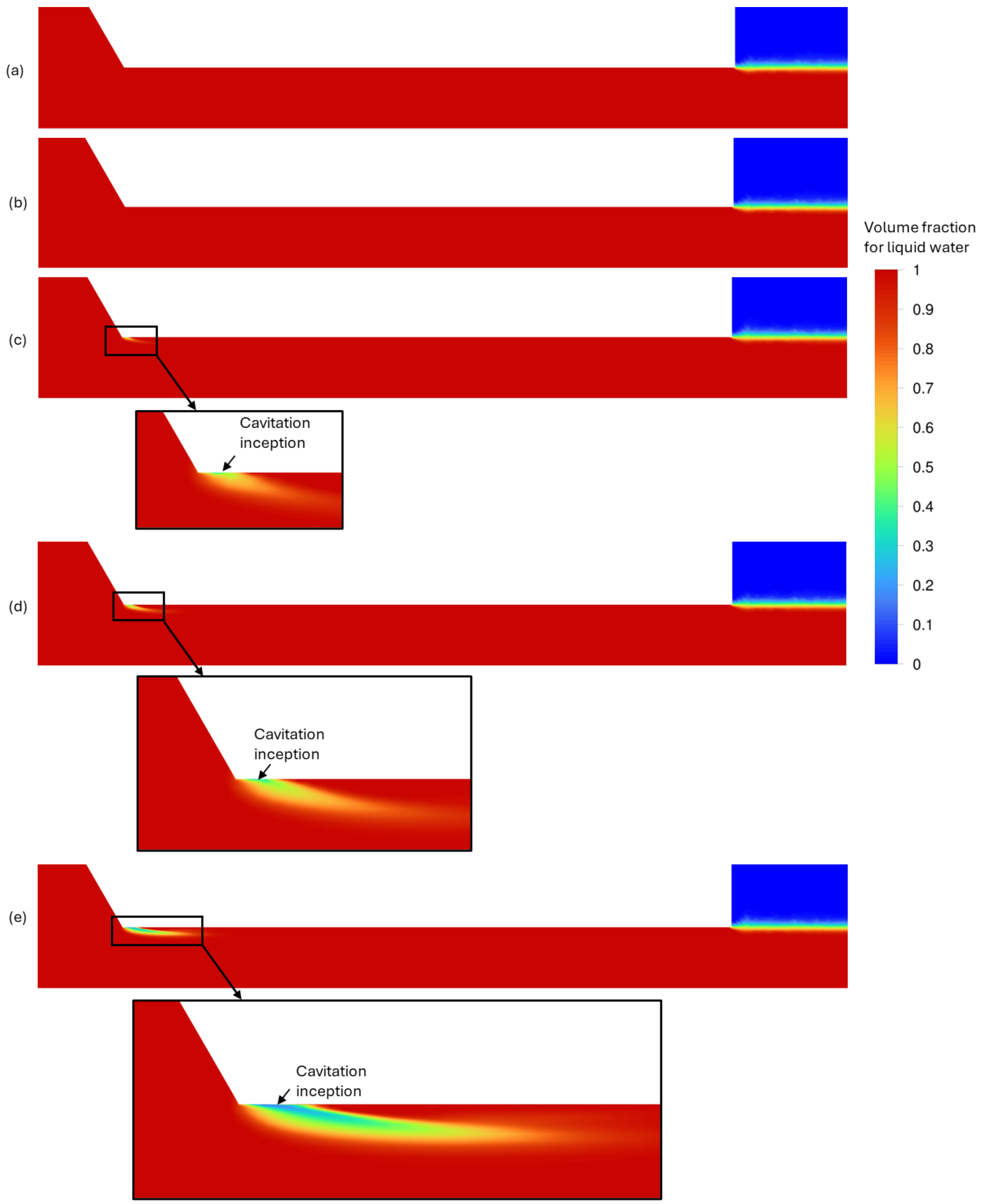

**Fig. 8 Contours of water (liquid) volume fraction for the chamfered orifice inlet and $L/D = 5$ and with (a) $\Delta p = 0.07$ MPa (b) $\Delta p = 0.14$ MPa (c) $\Delta p = 0.21$ MPa (d) $\Delta p = 0.27$ MPa, and (e) $\Delta p = 0.34$ MPa.**

## 3.3 Flow and cavitation results for the sharp-edge orifice inlet with $L/D = 2$, 5, and 10 for $\Delta p = 0.59$ MPa.

Table 4 compares the flow rates predicted by the current CFD study with the experimentally measured values by Kiaoulias et al (2019) for the sharp-edged orifice inlet at $\Delta p = 0.59$ MPa and $L/D = 2$, 5, and 10. The CFD-predicted flow rate matched closely with the experimental value for $L/D = 2$ (% difference of −1.3%). The % difference between CFD-predicted and experimental flow rates was 15.2% and 11.3% for $L/D = 5$ and 10, respectively.

**Table 4: CFD-predicted and experimental (Kiaoulias et al 2019) flow rates for the sharp-edge orifice inlet with $L/D = 2$, 5, and 10 for $\Delta p = 0.59$ MPa.**

| $L/D$ | $Q_{CFD}$(cc/min) | $Q_{Expt}$(cc/min) | % $Difference$ |
|---|---|---|---|
| 2 | 234.5 | 237.5 | −1.3% |
| 5 | 273.5 | 237.5 | 15.2 |
| 10 | 264.5 | 237.5 | 11.3% |

Fig. 9 shows the plots of liquid water volume fraction for the $L/D = 2$, 5, and 10 cases. A significant cavitation was observed for $L/D = 2$. However, a hydraulic flip, as postulated by Kiaoulias et al (2019) for the sharp-edged orifice inlet for $L/D = 2$, was not observed from the current CFD study. As mentioned in the section 3.2, a hydraulic flip was seen for $L/D = 5$ and that was consistent with the findings by Kiaoulias et al (2019). For $L/D = 10$, the cavitation zone extended to almost the entire orifice length, and a hydraulic flip was not observed. This is consistent with the hypothesis made by Kiaoulias et al (2019).

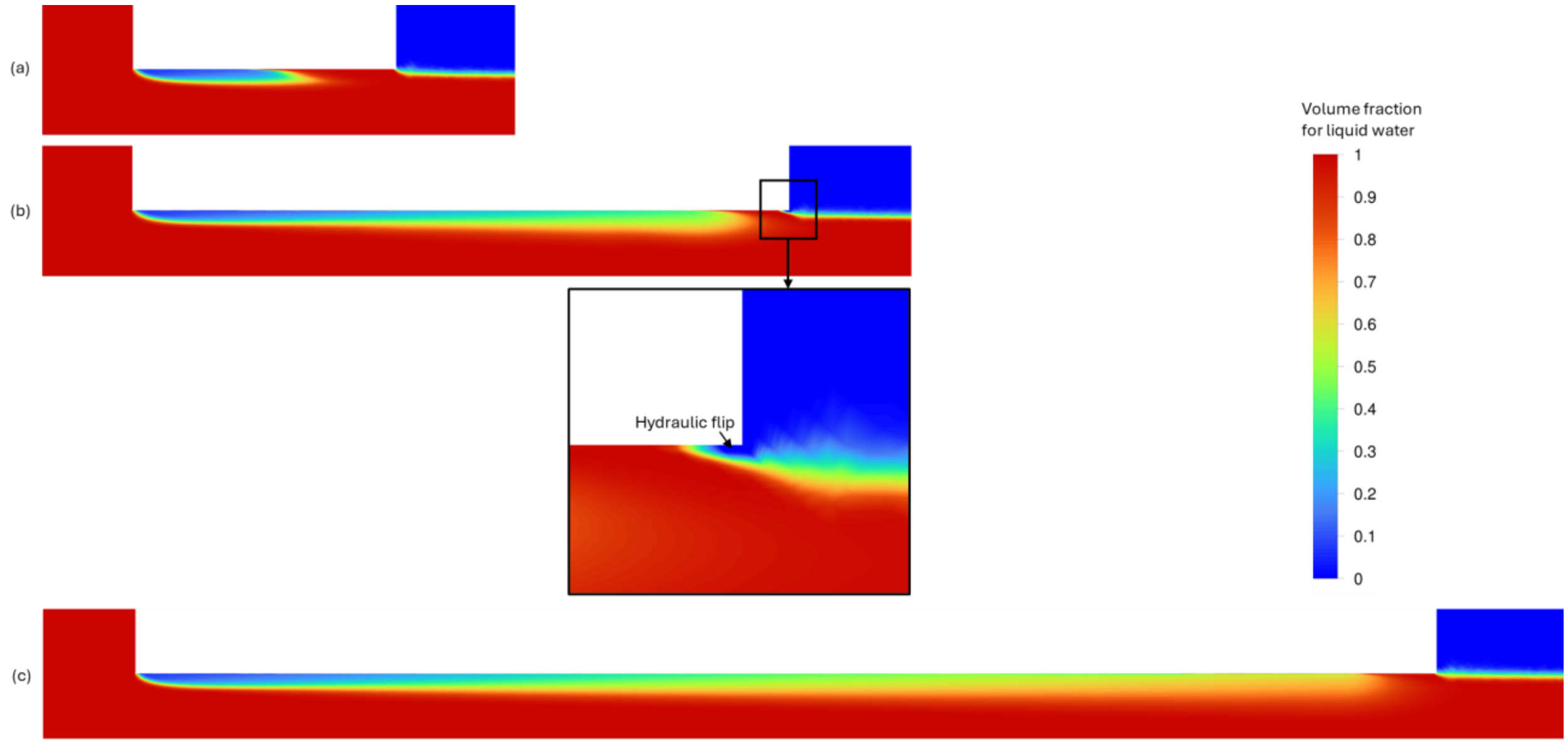

**Fig. 9 Contours of water (liquid) volume fraction for the sharp-edge orifice inlet with $L/D = 2$, 5, and 10 for $\Delta p = 0.59$ MPa.**

# 4. Conclusion

CFD simulations were carried out to predict the flow and cavitation in single-liquid injectors with sharp-edged and chamfered orifice inlets. The CFD-predicted volumetric flow rate was compared with the experimental values from previous work by Kiaoulias et al (2019). The CFD-predicted flow rate matched well with the experimental values for pressure drops lower than 0.2 MPa. For pressure drops larger than 0.2 MPa, the current CFD models overpredicted the flow rate with a peak % difference of 15% for the sharp-edged cases and a peak % difference of 9% for the chamfered cases. In summary, the flow prediction from the CFD model was validated against the previously published experimental values. Strong cavitation was observed for the sharp-edge orifice inlet with $L/D = 5$ at pressure drops larger than 0.2 MPa. Moreover, a hydraulic flip was observed for a pressure drop of 0.59 MPa. Cavitation inception was observed for the chamfered orifice inlet with $L/D = 5$ at pressure drops larger than 0.2 MPa. The cavitation results for the sharp-edged and chamfered orifice inlets for $L/D = 5$ were found to be consistent with the findings from the experimental study by Kiaoulias et al (2019). The $L/D$ of 2 and 10 were also simulated using the current CFD model. Although a significant cavitation was observed for these $L/D$, no hydraulic flip was observed. This observation for $L/D = 2$ was not consistent with the findings by Kiaoulias et al (2019). However, the CFD-based prediction of no hydraulic flip for $L/D = 10$ agreed with the findings by Kiaoulias et al (2019).

# Acknowledgement

This research was partially supported by Tier 1 Research Awards (ICDS_READ26_l1skg5951 and ICDS_READ26_l2skg5951) from the Institute for Computational and Data Sciences (ICDS) at The Pennsylvania State University. The authors are grateful to the ICDS for providing access to the Roar Collab Cluster. The authors acknowledge Dr. Grant Risha at Penn State Altoona for his consultation.

# Conflict of Interest

Authors have no conflicts to report.

# ORCID iD

Shreyash Gulati 0000-0001-6089-1421